\documentstyle{article} 
\begin{document}

\title{Parity-locking effect in a strongly-correlated ring}
\author{C.~A.~Stafford$^{1}$
and D.\ F.\ Wang$^2$\\
$\mbox{}^1${\small{\em Institut de Physique 
Th\'{e}orique, Universit\'{e} de Fribourg}}\\ {\small{\em 
CH-1700 Fribourg, Switzerland}}\\
$\mbox{}^2${\small{\em Institut de Physique Th\'eorique, EPF
Lausanne}}\\ {\small{\em 
CH-1015 Lausanne, Switzerland}}}
\maketitle

\begin{abstract}
Orbital magnetism in an integrable model of a multichannel
ring with long-ranged electron-electron interactions is investigated.
In a noninteracting multichannel
system, the response to an external magnetic flux is
the sum of many diamagnetic and paramagnetic contributions, but we find that
for sufficiently strong correlations, the contributions of all channels
add constructively, leading to a parity (diamagnetic or paramagnetic) which
depends only on the total number of electrons.
Numerical results confirm that this parity-locking effect is robust 
with respect to subband mixing due to disorder.
\end{abstract}

The free energy $F(\phi)$ of a metallic ring threaded by an Aharonov-Bohm flux
$(\hbar c/e)\phi$ is a periodic function of $\phi$, 
with period $2\pi$ \cite{byang}.  The system is said to be diamagnetic
if $F$ is minimal for $\phi=0$, and paramagnetic if $F$ is minimal
for $\phi=\pi$ (1/2 flux quantum).
A purely one-dimensional (1D) ring with $N$ spinless electrons 
is diamagnetic if $N$ is even and paramagnetic if $N$ is odd, 
independent of disorder and interactions \cite{leggett,loss}. 
This connection between the parity of $N$ and the sign of the magnetic
response of the system is sometimes referred to as {\em Leggett's theorem} 
\cite{leggett}.
The persistent current $I=-(e/\hbar)\partial F/\partial \phi$ 
is a periodic function of $\phi$ with amplitude $I_0=ev_F/L$
in a clean 1D ring \cite{bu}, where $v_F$ is the Fermi velocity and $L$ the 
circumference of the ring.  In a ring with many independent channels, 
$I$ is the sum of many such diamagnetic and paramagnetic contributions, 
and thus has a random sign and small amplitude \cite{byang}.  
Leggett's theorem is thus generally violated in multichannel systems.

In an interacting system, the parities of different channels are no
longer independent.  In this article, we investigate the orbital magnetism
of 1D $SU(M)$
fermions interacting via the potential $V(x)=g/x^2$. 
We find that for $g>0$, Leggett's theorem is restored for an arbitrary number
of channels $M$.  In addition, provided the ring is sufficiently thin
($k_F L > 2 \pi M$) we find that the 
magnetic response of all channels add constructively,
leading to a large enhancement of the persistent current.
A disordered two-channel ring with interchannel
interactions is also investigated numerically, and shows,
importantly, that the parity-locking effect persists even when the subbands are
mixed strongly by disorder.  
Interchannel correlations of this sort may be important to explain the 
anomalously large observed value of the persistent current in normal metal
rings \cite{levy,webb}.

Interacting spinless
electrons in a non-disordered ring with $M$ transverse channels,
threaded by an Aharanov-Bohm flux $(\hbar c/e)\phi$, may be represented by
1D $SU(M)$ fermions.  The transverse degrees
of freedom may be represented by an $SU(M)$ spin variable $\sigma=1,\ldots,M$. 
In the absence of
disorder, and for interactions which depend only on the electrons' 
coordinates along the ring ({\it i.e.}, for thin rings), the number of
electrons $K_{\sigma}$ in each channel is conserved.
The Hamiltonian of the system is 
\begin{equation}
H  =  -\frac{1}{2}\sum_{i=1}^{N} \frac{\partial^2}{\partial x_i^2} 
+ \sum_{i<j} V(x_i-x_j)
+ \sum_{\sigma=1}^{M} K_{\sigma} \varepsilon_{\sigma},
\label{ham.gen}
\end{equation}
where $N=\sum_{\sigma} K_{\sigma}$ is the total number of electrons and
$\varepsilon_{\sigma}$ is the energy minimum of subband $\sigma$.
Units with $\hbar=m=1$ are used.
The Aharanov-Bohm flux leads to the twisted boundary condition \cite{byang}
\begin{equation}
\Psi(x_1\sigma_1, \cdots, (x_i+L)\sigma_i, \cdots, x_{N}\sigma_{N})
= e^{i\phi}
 \Psi(x_1\sigma_1, \cdots, x_i\sigma_i, \cdots, x_{N}\sigma_{N}). 
\label{eq:boundary}
\end{equation}
For simplicity, let us consider equally spaced subbands $\varepsilon_{\sigma
+1} -\varepsilon_{\sigma}=\Delta\equiv E_F/M$.  The subband splitting $\Delta$
plays the role of an $SU(M)$ magnetic field.  
As we shall see, the effect of
repulsive interactions is to renormalize this effective field,
causing a condensation of electrons into the lowest subband.  
At $T=0$, the equilibrium persistent current is given by 
$I(\phi) = -(e/\hbar) \partial E_0/\partial \phi$, 
where $E_0(\phi)$ is the ground
state energy of Eq.\ (\ref{ham.gen}), subject to the boundary condition
(\ref{eq:boundary}).  

The first harmonic of the persistent current may be characterized by the
value of $I$ at $\phi=\pi/2$ (1/4 flux quantum), assuming the higher odd
harmonics are small.  For $V(x)=0$, one finds
\begin{equation}
I(\phi_0/4) = \frac{e\hbar \pi}{2 m L^2}
\sum_{\sigma=1}^M (-1)^{K_{\sigma}} K_{\sigma}.
\label{I0}
\end{equation}
For $M\gg 1$, this leads to the well-known \cite{Ityp} result $|I| \sim
M^{1/2} I_0$ due to the random parities of the different channels.  
The system may be either diamagnetic or paramagnetic, depending on the
channel occupancies $K_{\sigma}$.

Let us next consider a model with long-range interactions:
$V(x)=g/d(x)^2$, where $d(x)=(L/\pi)|\sin(\pi x/L)|$ is the chord length
along the ring.  This model was introduced and solved by Sutherland
\cite{sutherland} for the case $M=1$ and $\phi=0$.
First, we show that the $SU(M)$ model with twisted boundary conditions
is completely integrable.
Let us introduce the following generalized momentum operators
\begin{equation}
\pi_j=p_j - i \lambda
{\pi\over L} \sum_{k(\ne j)} \cot{(x_j-x_k)\pi\over L} P_{jk},
\end{equation}
where the operator $P_{jk}$ permutes the isospins and the positions of the
two particles $j$ and $k$ simultaneously, and $\lambda=\sqrt{g+1/4}+1/2$. 
These generalized 
momentum operators are hermitian, $\pi_i =\pi_i^\dagger$.  
The Hamiltonian  can be expressed as
a quadratic form in terms of the $\pi_i$,
\begin{equation}
H=\sum_{i=1}^N {1 \over 2} \pi_i^2 +\sum_{\sigma=1}^M \epsilon_\sigma K_\sigma
+\mbox{const}. 
\end{equation}
Introducing a set of new operators
$\tilde \pi_i=\pi_i + \lambda \sum_{j(\ne i)} P_{ij},$
one can show that the following infinite number of physical
operators commute with each other
\begin{equation}
[\tilde I_n, \tilde I_m]=0, 
\end{equation}
where $\tilde I_n= \sum_{i=1}^N (\tilde \pi_i)^n$, with $n=0,1,2,\ldots,\infty$.
Furthermore, one can show that these operators also commute with the 
Hamiltonian
\begin{equation}
[H, \tilde I_n]=0. 
\end{equation} 
This therefore provides a proof of the complete integrability of the system.
Our proof of integrability follows that given by 
Polychronakos for the Calogero-Sutherland model
\cite{poly}.  Note, however, that the operator $P_{ij}$ permutes the positions 
and the isospins of two particles $i$ and $j$, while 
the operator $M_{ij}$ of Ref.~\cite{poly} only exchanges the positions of 
the two particles.

Consider the following Jastrow wavefunction 
for given channel occupancies $K_{\sigma}$
\begin{equation}
\tilde \Psi_{0} (x_1\sigma_1,x_2\sigma_2,\cdots,x_N\sigma_N)
= \prod_{1\le i< j\le N}
|\sin({x_i-x_j\over L}\pi)|^{\lambda-1} \sin({x_i-x_j\over L}\pi),
\label{gsabs}
\end{equation}
where the isospins of the particles are symmetric under exchange.  
This wavefunction satisfies periodic boundary conditions ($\phi=0$)
if $N$ is odd and
satisfies antiperiodic boundary conditions ($\phi=\pi$) if $N$ is even.
It also satisfies Fermi-Dirac statistics.
Furthermore, this wavefunction is annihilated by 
the operators $\pi_i$, with $i=1,2,\cdots, N$:
\begin{equation}
\pi_i \tilde \Psi_{0} (x_1\sigma_1,x_2\sigma_2,\cdots,x_N\sigma_N)=0, 
\end{equation} 
which indicates that it is the ground state of $H$ in the subspace with fixed
$K_{\sigma}$.  The absolute ground state of the system thus occurs for
$\phi=0$ when $N$ is odd and for $\phi=\pi$ when $N$ is even, regardless of
the subband spacing $\varepsilon_{\sigma}$.  The Leggett theorem thus
holds for this interacting multichannel system, {\it i.e.}, the ground
state is diamagnetic if $N$ is odd and paramagnetic if $N$ is even.

Now, let us consider general values of the external magnetic flux.
For $g>0$, the ground state is highly degenerate
in the limit $L\rightarrow \infty$ in the absence of $SU(M)$ symmetry
breaking ($\Delta= 0$) due to the strong repulsion of the potential
at the origin, which prohibits particle exchange.  In a finite ring, 
one expects many $SU(M)$ level crossings \cite{period}
as a function of $\phi$. These
states differ in energy by at most $\pi \hbar v_F/L$ due to boundary effects;
all electrons will thus be condensed into the lowest subband for 
$\Delta>\pi\hbar v_F/L$, {\it i.e.}, for $k_F L > 2\pi M$, which is 
satisfied provided the ring is sufficiently thin.
The ground state of the system in this ``ferromagnetic'' state
has the Jastrow product form
\begin{equation}
\Psi (\{x\})
= \exp\left(i\frac{\phi-a}{L} \sum_{k=1}^N x_k\right) \prod_{1\le i< j\le N}
\left|\sin\left({x_i-x_j\over L}\pi\right)\right|^{\lambda-1} 
\sin\left({x_i-x_j\over L}\pi\right)
\end{equation}
for $0\leq \phi \leq \pi$, where
$a=0$ if $N$ is odd and $a=\pi$ if $N$ is even.
One readily verifies that $\Psi$ is an eigenstate of Eq.\ (\ref{ham.gen}),
has the correct symmetry, and obeys the twisted boundary condition 
(\ref{eq:boundary}).  
This state coincides with the absolute ground state $\tilde \Psi_0$ 
for $\phi=0\,(\pi)$ when $N$ is odd (even).
The ground state energy is found to be
\begin{equation}
E_0(\phi)=\frac{\pi^2 (\lambda+1)^2 N(N^2-1)}{6L^2} + 
\frac{N}{2} \left(\frac{\phi-a}{L}\right)^2.
\end{equation}
The corresponding persistent current is 
\begin{equation}
I(\phi_0/4)= (-1)^N \frac{e\hbar \pi N}{2 m L^2} 
\sim (-1)^N M I_0.
\label{Ilock}
\end{equation}
The condensation of all electrons into the lowest subband caused by
the strong repulsive interactions
thus leads to an enhancement of the typical persistent
current in the ballistic regime
by a factor of $M^{1/2}$ compared to that for noninteracting electrons,
given in Eq.\ (\ref{I0}). 

A peculiarity of the integrable model considered above is that the
number of electrons in each channel is a constant of the motion.  Both
disorder and more realistic interactions which depend on the transverse
coordinate will break this symmetry, and it is therefore important to 
verify that the parity-locking effect is not destroyed.  To this end, we
have considered a disordered two-channel ring, modeled in the tight-binding
approximation, with a nearest-neighbor interchain interaction $V$ included to 
induce interchannel correlations.
The Hamiltonian is
\begin{equation}
H= \sum_{i=1}^L \left[\sum_{\alpha=1}^2 
\left(e^{i\phi/L} c_{i\alpha}^{\dagger} c_{i+1 \alpha}
+ \mbox{H.c.}+\varepsilon_{i\alpha}\rho_{i\alpha}\right) + \frac{\Delta}{2}
\left(c_{i1}^{\dagger} c_{i2} + \mbox{H.c.}\right)
+V\rho_{i1}\rho_{i2}\right],
\label{ham.num}
\end{equation}
where $c_{i\alpha}^{\dagger}$ creates a spinless electron at site $i$ of chain
$\alpha$, $\rho_{i\alpha}\equiv c_{i\alpha}^{\dagger}c_{i\alpha}$, 
and $\varepsilon_{i\alpha}$ is a random
number in the interval $[-\varepsilon/2,\varepsilon/2]$.  
The interchain hopping determines the subband splitting $\Delta$.
Fig.\ 1 shows the persistent current for rings with 5 spinless
electrons on 18 sites  
as a function of $V$, calculated using the Lanczos technique.
Both the ensemble average of $I$ for 500 systems (squares) and the
persistent current of a typical system (solid curve) are indicated.
The error bars indicate the standard deviation $\delta I =
(\langle I^2\rangle-\langle I\rangle^2)^{1/2}$ over the ensemble (not the
statistical uncertainty in the mean value).
The subband splitting $\Delta=0.8$
is chosen so that in the absence of disorder and interactions, $K_1=3$ and
$K_2=2$, leading to a large cancellation of the persistent current due to
the different parities of the two channels.
The on-site disorder $\varepsilon=2>\Delta,\,E_F$ mixes the two
channels, but does not lead to strong backscattering.
For $V=0$, the sign of $I$ is essentially random, and $\langle I \rangle$
is slightly diamagnetic.  As $V$ is increased, $\langle I \rangle$
oscillates in sign and increases in magnitude, becoming strongly diamagnetic
for large $V$.  $|\langle I\rangle|$ is increased by a factor of 10 as
$V$ is increased from 0 to 20, while $\sqrt{\langle I^2\rangle}$ is 
increased by a factor of 3.  The system exhibits parity locking for large
$V$.

While the subband occupancies are no longer constants of the motion in
Eq.\ (\ref{ham.num}),
there is a corresponding
topological invariant in the disordered system, namely,
the number of transverse nodes in the many-body wavefunction \cite{leggett}
({\it i.e.},
nodes which encircle the AB flux $\phi$).  The lowest subband has no
such nodes, while each electron in the second subband contributes one 
transverse node.  In order for two electrons to pass each other as they
circle the ring, such a transverse node must be present.  As $V$ increases,
it becomes energetically unfavorable for electrons to approach each
other, so transverse nodes in the many-body wavefunction will tend to
be suppressed.  In the strongly-correlated limit, all such nodes will be
eliminated, leading to a state whose parity depends only on the {\em total}
number of electrons. 
In such a state, the persistent currents
of all channels add constructively, leading to a large persistent
current (see Fig.\ 1).  It should be emphasized that while the enhancement
of the persistent current shown in Fig.\ 1 is relatively modest, a
much larger enhancement would be expected in a system with $M\gg 1$, based
on the above arguments.

In conclusion, it has been shown that for sufficiently strong electronic
correlations, the sign of the orbital magnetic response in a multichannel
ring is determined solely by the parity of the total number of electrons in the
system.  The magnetic response of all channels add constructively, leading
to a large enhancement of the persistent current.
This result, which we refer to as the parity-locking effect,
was demonstrated in an integrable model with long-range
interactions as well as in a disordered two-channel system.
It should be emphasized that the parity-locking effect holds
in both ballistic and disordered systems; it is therefore
complementary to mechanisms previously proposed for the enhancement of the
persistent current \cite{eckern,screening,screening2,1D-analytic,2D-numerics}, 
which rely on the competition between disorder
and interactions.  It is likely that both such mechanisms are important to
explain the anomalously large observed value \cite{levy,webb}
of the persistent current in disordered metallic rings.

C.\ A.\ S.\ thanks M.\ B\"uttiker for valuable discussions.
This work was supported by the Swiss National Science Foundation.

\begin{figure}
\vbox to 14cm {\vss\hbox to 17cm
  {\hss\
    {\includegraphics{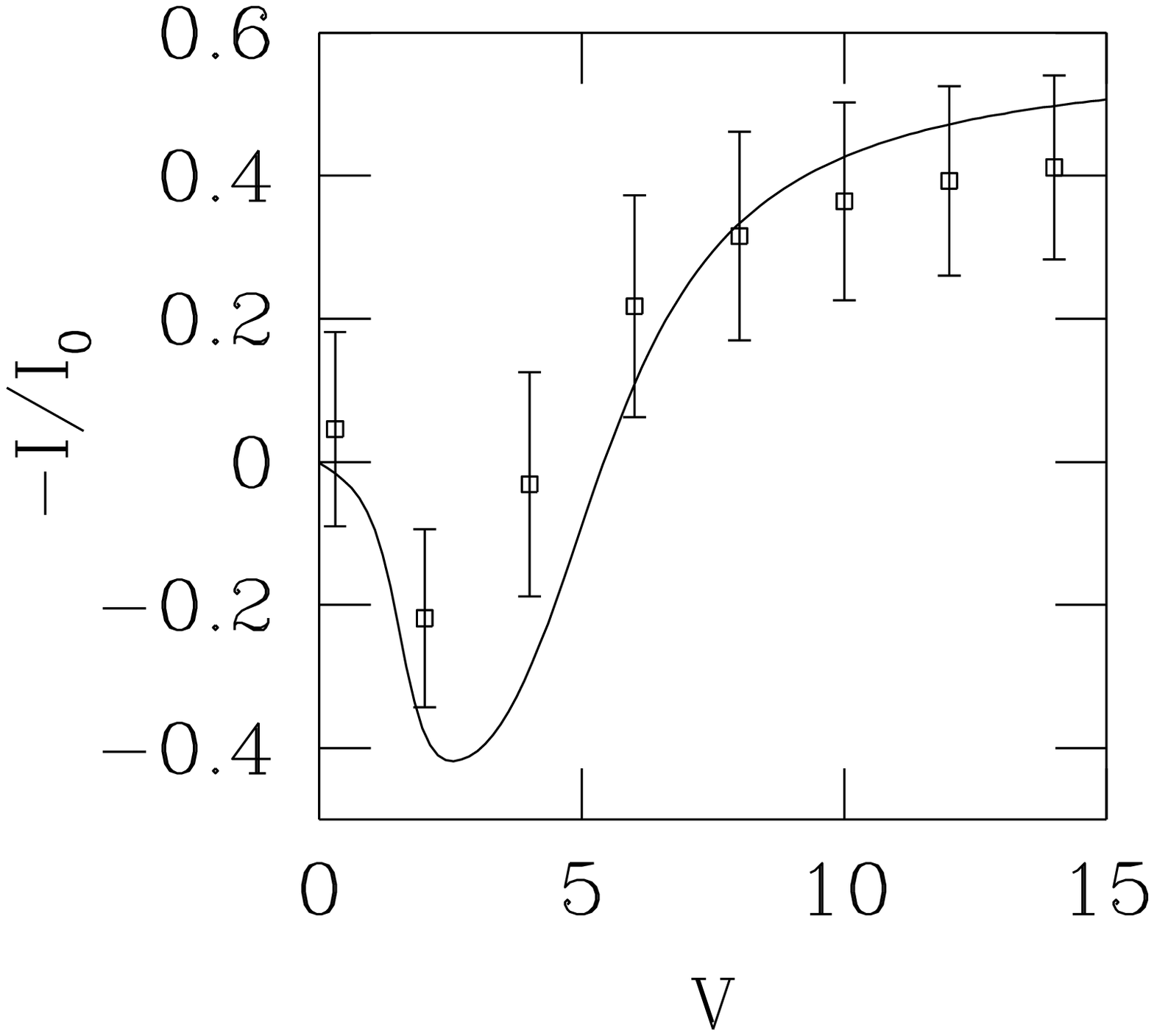}}
   \hss}
}
\caption{Persistent current $I=-(e/\hbar)\partial E_0/\partial 
\phi|_{\phi=\phi_0/4}$ of a disordered two-channel ring with 5 spinless
electrons on 18 sites as a function of the interchain interaction $V$. 
The current is given in units of $I_0=ev_F/L$.
The amplitude of the on-site disorder is $\varepsilon =2$ and the
subband splitting is
$\Delta=0.8$.  Solid curve: persistent current for one
realization of disorder.  Squares: ensemble average $\langle I\rangle$
for 500 systems.  The error bars indicate the width $\delta I=
(\langle I^2\rangle -\langle I\rangle^2)^{1/2}$ of the current distribution.
Note that the persistent current is diamagnetic for
large $V$, as expected for a system with $N$ odd due to parity-locking; 
this is true for all realizations of disorder.}
\label{system}
\end{figure}

\end{document}